\documentclass[11pt]{article}

\usepackage{graphicx}
\def \b {\begin{eqnarray}}
\def \el#1 {\label{#1} \e}
\def \e {\end{eqnarray}}
\def \ds {\displaystyle }

\newcommand{\ave}[1]{\langle {#1} \rangle}

\begin{document}

\title{A new view of the spin echo diffusive diffraction on porous structures}
\date{March 2002}
 \author{\thanks{Physics Department, FMF,
University of Ljubljana, Jadranska 19, 1000
Ljubljana, e-mail: Janez.Stepisnik@fiz.uni-lj.si,  Slovenia} Janez Stepi{\v s}nik}

\maketitle

\begin{abstract}
Analysis with the characteristic functional of stochastic motion is used for the gradient spin echo measurement of restricted motion to clarify details of the diffraction-like effect in a porous structure.  It gives the diffusive diffraction as an interference of spin phase shifts due to the back-flow of spins bouncing at the boundaries, when mean displacement of scattered spins is equal to the spin phase grating prepared by applied magnetic field gradients. The diffraction patterns convey information about morphology of the surrounding media at times long enough that opposite boundaries are restricting  displacements.  The method explains the dependence of diffraction on the time and width of gradient pulses, as observed at the experiments and the simulations. It also enlightens the analysis of transport properties by the spin echo, particularly in systems, where the motion is restricted by structure or configuration.
\end{abstract}
\newpage
\section{Introduction}

Measurements of the molecular displacements through precession of their atomic nuclear spins in non-uniform magnetic field by the magnetic resonance spin echo\cite{Hahn} has gained a most decisive role at the studies of molecular transport within porous structures that may comprise system as diverse as sandstone rocks, catalysts, colloids, or biological tissue. The field has an extensive literature and a wide range of experiments has been performed. The methodology of the Pulsed Gradient Spin Echo (PGSE)~\cite{Stejskal651} has been successfully implemented to measure diffusion in systems for which a constrained molecular motion caused a deviation from Fickian behavior.  Based on the diffusion propagator formalism~\cite{Karger}, which gives the spin echo as a Fourier transform of the probability distribution~\cite{Cory}, Callaghan has introduced the concept of diffusive diffraction of spin echo in a porous media~\cite{Callaghan}. In this way, one can extract information not only about the motion but also about the morphology of the surrounding medium~\cite{Callaghan91,Kuchel,Seymour,Panfilis}. Unlike NMR Imaging where molecular position are recorded to a resolution on the order of $10\mu m$, the method is able to achieve the resolution of displacement measurement some two to three orders of magnitude better, and pushes the lower limit of NMR resolution into a nanometer range. 

However, the use of probability distribution function is just one of ways to consider the transport properties of spins in terms of the averaged characteristics of the elementary events. The method of characteristic functional, {\it i.e.} the Fourier transform of the probability distribution~\cite{Kampen}, is an alternative way to average the spin phase fluctuation. With the cumulant expansion in the Gaussian approximation, it gives the relation between the spin echo and the spectrum of the single particle velocity correlation function (VCF)~\cite{moj81,moj85,moj981}. It provides a simple derivation of spin echo attenuation for any gradient pulse sequence. It also shows that a sufficiently fast and a properly shaped gradient sequence, like the Modulated Gradient Spin Echo (MGSE)~\cite{mojcall3,mojcall}, conveys information not only about macroscopic flow and diffusion, but also about the details of motion on the molecular level in the frequency range of gradient modulation. The method has been successfully implemented to measure the diffusion spectrum and the flow dispersion in porous media\cite{moj001,moj201}.

This study is an attempt to apply this method to the analysis of the spin echo diffusive diffraction. It aims to clear up details like the diffraction dependences on time, the dependence on the width of gradient pulse and on the type of applied gradient sequences as obtained by experiments and computer simulations~\cite{Callaghan92,Soderman,Blees}, but remains unanswered in the frame of propagator theory. The method might elucidate the dependence of diffraction on the mechanism of scattering at interfaces as well.

\section{Spin echo and average of spin phase fluctuation}

Whenever in NMR a non-uniform magnetic field is used to encode the spin magnetization for motion rather than position, the spin echo is used to refocus any spin phase shift, due to absolute spin position. Thus, the perturbations of spin phase, due to displacements ${\bf r}(t)$ of spins in the non-uniform magnetic field  gradient ${\bf G}(t)$, can be written as $\theta(\tau)=\gamma\int_0^{\tau}\,{\bf G}(t)\cdot {\bf r}(t)\,dt=-\int_0^{\tau}\,{\bf F}(t)\cdot {\bf v}(t)\,dt$, where ${\bf F}(t)=\gamma\int_0^t {\bf G}(t')dt'$ is a factor of spin dephasing, $\tau$ is the time of refocusing, and  ${\bf v}(t)$ is spin velocity.  Regarding their location in a non-uniform magnetic field and rf exitation, one can distinguish subgroups of spins according to their precession frequency to obtain the spin echo as~\cite{moj993}
\b
E(\tau)&=&\sum_j E_{jo} \ave{e^{\ds{- i\int_0^{\tau}{\bf
F}_j(t) \cdot {\bf v}_{j}(t)\,dt }}},
\el{spin_echo_sub}
where $E_{jo}$ is the normalized amplitude and the the motion average of particles in the $j$-th sub-ensemble is $\ave{ ... }$ t.  In the case of a sequence with two sharp gradient pulses of widths $\delta$ and interspaced for $\Delta$, {\it i.e.} a sharp PGSE sequence where ${\bf F}={\bf q}=\gamma\delta{\bf G}$, the mean of the spin phase fluctuation can be worked out with  the conditional probability distribution (the diffusion propagator) $P({\bf r'},t'|{\bf r},t)$, ~\cite{Karger,Cory,Callaghan}. It gives  the spin echo as
\b
E(\tau,{\bf q})&=&\sum_j E_{jo}\ave{ e^{\ds{i{\bf q}\cdot( {\bf  r}_j(\tau)-{\bf  r}_j(0)}}}\rightarrow \int\,E({\bf r'})d{\bf r'}\int\,d{\bf r}P({\bf r},\tau|{\bf r'},\Delta)e^{\ds{i{\bf q}\cdot( {\bf  r}-{\bf  r'})}}.
 \el{propagator}
Clearly, the spin echo appears as a Fourier transform of the probability distribution with respect to  ${\bf q}$~\cite{Cory}. Thus, the measurement of $E(\tau,{\bf q})$ gives an average propagator of restricted diffusion. This approach also explains the diffusive diffractions, which appear at long diffusion times and strong gradients~\cite{Callaghan}. 

According to statistical physics, the stochastic process is determined knowing either the probability distribution or the characteristic function, which is the Fourier transform \cite{Kubo2} of probability distribution, 
\b
&\ave{ e^{\displaystyle{ i\,f\,r}}}=\int e^{\ds{i\,f
\,r}}P(r)\,d{r},
\el{fcfunction}
The characteristic function exists even when the probability does not. By passing the stochastic properties from  $r$ to  $v$, where $r=\ds{\int_0^t v(t')dt'}$, the characteristic function is transformed into the characteristic functional that can be  worked out with the cumulant expansion method as
\b
 \ave{ e^{\ds{ i\int_0^{t}\,f(t') v(t')\,dt' }}},
&=& e^{\ds{ i\int_0^{t}\,f(t')
 \ave{v(t')} \,dt'-{1\over 2}\int_0^tdt_1\int_0^tdt_2\,f(t_1).\ave{ v(t_1)
 v(t_2)}_cf(t_2)+...}},
 \e
where $f(t)$ is an arbitrary function. With the cumulant expansion of characteristic functional, the stochastic process is determined by the expectation $\ave{v(t)}$, and by the set of correlation functions, of which the second order is $\ave{v(t_1) v(t_2)}_c=\ave{v(t) v(0)}-\ave{v(t)}^2$.  Gaussian approximation of the stochastic process is the expansion to the second term, which is often used to describe physical processes and can be justified in many processes in the magnetic resonance as well.  
 
These basic facts from the theory of stochastic processes leads to understanding that the average of the spin phase fluctuation can be resolved in different manners depending on the applied gradient sequence. With a sharp PGSE sequence, the position of spin-bearing particle ${\bf r}(t)$ is considered as a stochastic variable permitting to treat the spin phase fluctuation by using a probability distribution as shown in Eq.\ref{propagator}. In the cases of finite gradient pulses, multi-pulse gradient sequences or gradients of general waveform, the short gradient pulse approximation fails, and the stochastic properties can be passed to the velocity of spin, ${\bf v}$. As the spin phase average has a form of characteristic functional, it can be treated by the well-developed methods of statistical physics. With the cumulant expansion in the Gaussian approximation, the spin echo appears as
\b
 {E(\tau )} =\sum_j{E_{jo}\,e^{\ds i\,\phi_j(\tau)-\beta_j(\tau)}}.
\el{dump}
where the phase shift depends on the mean local spin velocity  
\b
\phi_j(\tau)=-\int_0^{\tau}{\bf F}(t)\ave{{\bf v}_j(t)}dt,
\el{phase}
and the spin echo attenuation is related to the velocity correlation function  
\b
\beta_j(\tau)={1\over
2}\int_0^{\tau}\int_0^{\tau} {\bf  F}_j (t_1)\cdot\ave{{\bf
v}_j(t_1)\,\,{\bf  v}_j (t_2) }_c \cdot{\bf F}_j(t_2) \,dt_1dt_2.
\el{att}

For a free motion in a simple fluid, the mean velocity is assumed to be zero $\ave{{v}_{gj}(t)}=0$, while VCF ca be approximated with $\ave{{v}_{gj}(t)\,\,{v}_{gj} (0) }=2\,D\,\delta(t)$, due to a short memory of molecular collisions. Substitutions into Eq.\ref{dump} gives a spin echo with exponential decay, which is in proportion to the mean squared displacement. According to Eq.\ref{fcfunction}, its  Fourier transform, with respect to parameter $\int_0^{\tau}{\bf F}(t)^2\,dt$, gives the probability distribution function for a free Fick's diffusion as expected.

However, for motion in complex systems, there are a number of characteristic time-scales, which may correspond to frequencies in the time regime of spin echo. These include a long tail decay of VCF in liquids\cite{Alder,Dong}, tube disengagement times in entangled polymers\cite{Schweizer}, a characteristic negative decay of VCF in confined fluids\cite{Hagen,Oppenheim}. Such times are more closely related to the structural dynamics of a liquid than to local particle motion. It usually appears as an anomalous time dependent attenuation of the spin echo.

In the previous implementations of cumulant expansion method in the analysis of the spin echo measurement of restricted motions~\cite{ moj93,moj981}, the cumulants were averaged over the volume of confinement. Therefore, the phase shift of spin echo has been neglected because of $\overline{\ave{\bf v}}=0$,and all attention was focused to the second term of expansion. It provides a known relation between the spin echo attenuation and VCF~\cite{moj201}. Such approximation is permitted as long as the gradient sequences are short enough or the gradients are weak enough that the spin displacement is short compare to the spin phase grating, $\ave{({\bf F}.\Delta {\bf r})^2}\ll 1$. Whenever it is not fulfilled, local distributions of the phase shifts and of the spin echo attenuation must be carefully considered. 

\section{Distribution of velocity correlation function and mean velocity of confined motion}

For the diffusion in restricted geometries, the approximations with $\ave{{v}_{g}(t)}=0$ and with VCF as a delta function are reasonable as long as the number of molecular impacts at walls is small compare to the number of intermolecular collisions, $\tau_c\ll t\ll\tau_w$. At longer times, we need better approximations. The solution of the Langevin equation for diffusion between parallel planes provides such VCF~\cite{Oppenheim}, but an identical results one can obtain with the use of the probability distribution from the Fick's diffusion equation to average velocity correlations as shown in reference~\cite{Duh2}. Therefore, we use this method to obtain the distribution of the mean velocity and VCF within the space of pore as follows. 

In the general form, the conditional probability function for a restricted diffusion is $$P({\bf r},t|{\bf r_o},t_o)=\sum_k\psi_k({\bf r})\psi_k({\bf r}_o)e^{-k^2\,D|t-t_o|}$$, where $ k$ and eigen functions $\psi_k({\bf r})$ characterize the compartment geometry. The mean spin velocity follows from the first derivative of the mean spin displacement as
\b
\ave{{\bf g}.{\bf v}({\bf r},t)}&=&\frac{d}{dt}\int_V{\bf g}.({\bf r}-{\bf r'})P({\bf r},t|{\bf r'})\,d{\bf r'}, 
\el{meanv}
where the unit vector $\bf g$ is aligned along the magnetic field gradient. The second derivative of the mean squared displacement gives VCF as
\b
\ave{{\bf g}.{\bf v}({\bf r},t)\,\, {\bf v}({\bf r},t).{\bf g}}&=&\frac{1}{2}\frac{d^2}{dt^2}\int_V({\bf g}.({\bf r}-{\bf r'}))^2P({\bf r},t|{\bf r'})\,d{\bf r'}
\el{VCF}

\begin{figure}[h]  
\centering \scalebox{0.8}{\includegraphics{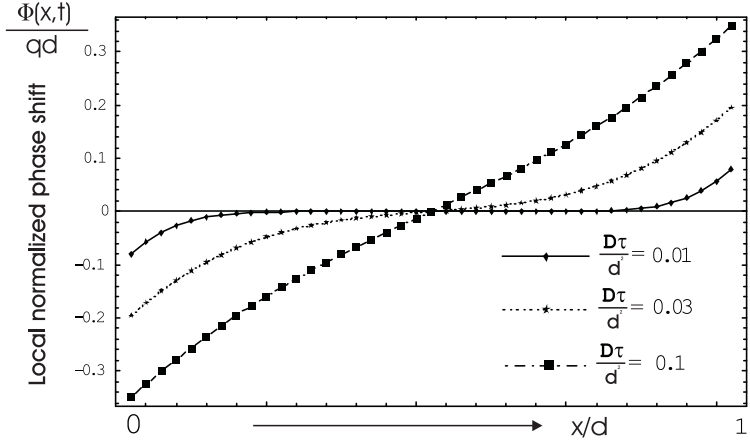}} \caption{Distribution of the diffusive phase shifts between plan-parallel planes for the sharp PGSE}\label{fig1} 
\end{figure}

Provided distributions make possible to obtain the spin echo for any gradient sequence by using relations of Eq.\ref{phase} and Eq.\ref{att}. Here, we consider the simplest case, {\it i.e.} the application of the sharp PGSE sequence  in order to enlighten the basic fact of diffusive diffractions from the point of view of new approach. With the eigen functions for the diffusion between plan parallel planes, Eq.\ref{phase}, Eq.\ref{att} with Eq.\ref{meanv} and Eq.\ref{VCF} give the distributions for the phase shifts and the spin echo attenuation at different times as shown in Fig.\ref{fig1}. At early times after the first gradient pulse, only spins in the proximity of wall are involved in the scattering. The component of mean velocity outward of boundaries gives the resulting phase shift that is in proportion to spin displacement, $\phi(\tau, {\bf r})\approx q\sqrt{D\,t }$, but with the opposite sign at the facing planes.  At longer time, when a number of scattered spins increases, the phase shift develops into almost linear dependence on location, $\phi(\tau, {\bf r})\approx {\bf q}({\bf r}-\ave{{\bf r}})$. 
\begin{figure}[h]  
\centering \scalebox{0.8}{\includegraphics{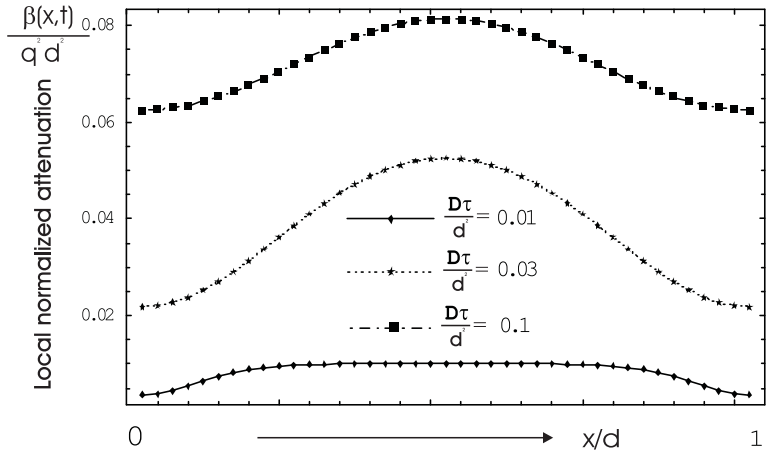}} \caption{Distribution of the diffusion attenuation between plan-parallel planes for the sharp PGSE}\label{fig2}
\end{figure}
As displayed in Fig.\ref{fig2}, the distribution of the spin echo attenuation at early times is almost uniform but with small dales in the proximities of the walls. With time, it develops into a well-defined distribution with a maximum in the center of compartment. 

With such strong dependences on the location, the averaging over the space of compartment for the mean velocity and VCF is not permitted. It requires the analysis that includes the local details of motion. Thus, the spin echo in the continuum limit has to be written  
\b
{E(\tau )} =\int_V{E({\bf r})\,e^{\ds i\,\phi(\tau, {\bf r})-\beta(\tau,{\bf r})}}\,d{\bf r}^3,
\el{distriecho}
where $\phi(\tau, {\bf r})$ and $\beta(\tau,{\bf r})$ describe the distribution of spin phase and of attenuation, respectively. 

\begin{figure}[h]  
\centering \scalebox{0.8}{\includegraphics{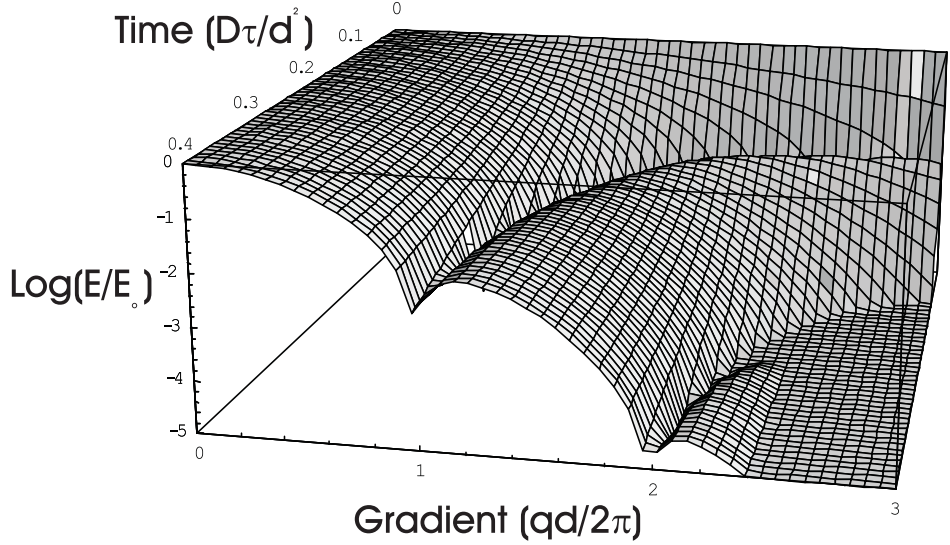}} \caption{Diffraction-like features of sharp PGSE as a function of time and gradient for diffusion between plan-parallel planes.}\label{fig3} 
\end{figure}
Integration of Eq.\ref{distriecho}, for the case of diffusion between parallel planes and with a sharp PGSE sequence, gives diffraction-like patterns as shown in Fig.\ref{fig3}. Diffractions exhibit a dependence on gradient magnitude $q$ as well as on time $\tau$. At short times, the diffraction minima are shifted toward larger $q$, and depend on the spin displacement as $2\sqrt{D\tau}q\approx n 2\pi$. At displacements large enough that a spin is starting to experience scattering on the opposite boundaries, the minima appear at value of $q\,a\approx n 2\,\pi$ where $a$ is about the diameter of pore. It occurs at the displacements above $2D\tau>0.3 a^2$ according to Fig.\ref{fig3}. At these times, the sharp PGSE obtains the form 
\b
E(\tau)\approx e^{\ds{-\frac{1}{2}{\bf q}^2(M_2-{\bf M}_1^2)}}|\int_Ve^{\ds{i{\bf q}.({\bf r}-{\bf M}_1)}}d{\bf r}|
\e
where ${\bf M}_1=\int_V\,{\bf r}\,d{\bf r}$ and $M_2=\int_V\,{\bf r}^2\,d{\bf r}$, and the diffractions can pass on an information about morphology.

For the diffusion between parallel planes that are interspaced for $a$, it gives 
\b
E(\tau)\approx e^{\ds{-\frac{1}{24}(qa)^2}}|\frac{2\sin(qa/2)}{q }|,
\el{plandiffr}
which becomes is in the limit of of small $q$ 
\b
E(\tau)\approx e^{\ds{-\frac{1}{12}(qa)^2}}.
\e
Although the last result is the same as that obtained with the propagator approach\cite{Callaghan}, the dependence of the spin echo diffraction patterns on $q$ as well as on the interval of measurement are very different as shown in Fig.\ref{fig3}. 

\section{Conclusion} 
The analysis with the cumulant expansion of the characteristic functional of stochastic motion provides some new details of diffraction-like features that occur at the gradient spin echo measurement of diffusion and flow within porous media. As in optics, the diffraction is interference of waves with different wave vector, here it is the interference of phase shifts because of back-flow of spins scattered at boundaries, whenever the repulsive displacements are comparable to the spin phase grating created by applied magnetic field gradients. At long times, when a spin starts to experience scattering on the opposite boundaries, the diffraction conveys information about morphology of the surrounding media. This approach is able to explain the diffraction dependence on the time and the duration of gradient pulses, as observed at experiments and simulations and can be implemented for any gradient sequence. It also casts a new light to the spin echo measurements of transport properties, particularly in complex systems where the motion is constrained by structure or configuration.

\section*{Acknowledgment}
I am grateful to the Slovenian  Ministry of Education, Science and Sport for the financial support.

\end{document}